\documentclass[10pt,twocolumn,twoside]{IEEEtran}
\ifCLASSINFOpdf
\else
\fi
\usepackage{amsmath}
\usepackage{amsthm}
\usepackage{amsfonts}
\usepackage{algorithm}
\usepackage{graphicx}
\usepackage{algorithmic}
\usepackage{amsthm}
\usepackage{hyperref}
\usepackage[noadjust]{cite}
\usepackage{color}
\usepackage{caption}
\usepackage{amssymb}
\usepackage{mathtools}
\usepackage{psfrag}

\begin{document}
\title{General Rank Multiuser Downlink Beamforming With Shaping Constraints Using Real-valued OSTBC}

\author{Ka Lung Law,
\thanks{

The authors are with the Communication Systems Group, Technische Universit\"at Darmstadt, D-64283 Darmstadt, Germany (e-mail: klaw@nt.tu-darmstadt.de; xwen@nt.tu-darmstadt.de; mpesa@nt.tu-darmstadt.de).}
Xin Wen,
Minh Thanh Vu,
and Marius Pesavento, {\it Member, IEEE}
}

\maketitle

\begin{abstract}
In this paper we consider optimal multiuser downlink beamforming in the presence of a massive number of arbitrary quadratic shaping constraints. We combine beamforming with full-rate high dimensional real-valued orthogonal space time block coding (OSTBC) to increase the number of beamforming weight vectors and associated degrees of freedom in the beamformer design. The original multi-constraint beamforming problem is converted into a convex optimization problem using semidefinite relaxation (SDR) which can be solved efficiently. In contrast to conventional (rank-one) beamforming approaches in which an optimal beamforming solution can be obtained only when the SDR solution (after rank reduction) exhibits the rank-one property, in our approach optimality is guaranteed when a rank of eight is not exceeded. We show that our approach can incorporate up to 79 additional shaping constraints for which an optimal beamforming solution
is guaranteed as compared to a maximum of two additional constraints that bound the conventional rank-one downlink beamforming designs. Simulation results demonstrate the flexibility of our proposed beamformer design.

\end{abstract}
\IEEEpeerreviewmaketitle

\section{Introduction}

With the massive growth of the number of wireless communication users and the increasing demands for high-rate services, the spectral resource is becoming more and more scarce. Research on spectrally efficient transmission schemes for current and next generation cellular networks that are capable of mitigating effects of multiuser and co-channel interference is attracting considerable interest \cite{dahlman20114g}. As a spectrally efficient multi-antenna technique \cite{wirelessbook}, downlink beamforming has been extensively studied in the past few years \cite{rashid1998, bengtsson1999optimal, bengtsson2001optimal, schubert2004solution, Shamai2006, gershman2010convex}. With the aid of channel state information (CSI) at the transmitter, downlink beamforming is employed at the base station of cellular networks to serve multiple co-channel users simultaneously using spatially selective transmission.

As a pioneering work in downlink beamforming, the authors in \cite{rashid1998} consider the problem of minimizing the total transmitted power subject to quality of service (QoS) constraints in the form of minimum signal to interference plus noise ratio (SINR) requirements at each user. A particular form of uplink-downlink duality theory is established in \cite{rashid1998} and under this framework the downlink beamforming problem is solved using a computationally efficient power iteration algorithm. A similar approach that exploits uplink-downlink duality is proposed in \cite{schubert2004solution}, where the downlink beamforming problem of maximizing the minimum SINR subject to a total power constraint is considered.

A different class of approaches is presented in \cite{bengtsson1999optimal, bengtsson2001optimal} and \cite{Shamai2006} where the downlink beamforming problem is addressed using conic optimization. The authors in \cite{bengtsson1999optimal, bengtsson2001optimal} solve the beamforming problem by resorting to the concept of semidefinite relaxation (SDR) and prove that from the rank-relaxed problem a rank-one solution\footnote{By a rank-one solution we mean that the solution matrices of the SDR problem exhibits the rank-one property.} can always be obtained when the problem is feasible. Besides, the authors in \cite{Shamai2006} cast the problem into a computationally efficient standard second order cone program (SOCP) and for which the corresponding optimality conditions are derived.

The multiuser downlink beamforming approaches referenced above all optimize the beamforming weights considering the SINR requirements of the individual users served in the network. In addition to this, supplementary shaping constraints on the beamforming weight vectors are embedded in the downlink beamforming problem to incorporate a variety of requirements in diverse applications \cite{cognitiveradio1, cognitiveradio2, cognitiveradio3, femtocell1, femtocell2, femtocell3, physicalsecurity1, physicalsecurity2, physicalsecurity3, wirelesscharging1, wirelesscharging2, wirelesscharging3, hammarwall2006,hammarwall2005}. For example, in hierarchical cellular networks operating under the licensed shared access (LSA) paradigm \cite{LSA}, pico- and femtocell networks co-exist in the same frequency band with the surrounding macrocell \cite{femtocell1, femtocell2, femtocell3}. Shaping constraints are employed at the femtocell base stations to limit the power leakage to the macrocell users \cite{cognitiveradio1, cognitiveradio2, cognitiveradio3} and the power leakage to concurrent femtocell networks \cite{femtocell1, femtocell2, femtocell3}. Similarly, in the newly emerging context of physical layer secrecy, shaping constraints are applied to guarantee that the SINRs at the eavesdroppers reside below a given detection threshold such that the confidential information can only be decoded at the desired receiver \cite{physicalsecurity1, physicalsecurity2, physicalsecurity3}. Recently downlink beamforming has been associated with wireless charging in energy harvesting communication networks. In this context shaping constraints are employed to ensure that the received power at the harvesting nodes is greater than a prescribed threshold to facilitate efficient charging \cite{wirelesscharging1, wirelesscharging2, wirelesscharging3}. Furthermore, shaping constraints are used in multiuser downlink networks to limit the interference power leakage to co-channel users, e.g., in neighboring cells \cite{hammarwall2006,hammarwall2005}.

The above mentioned SDR approach lends itself for application in the multi-constraint downlink beamforming problems with a large number of additional shaping constraints \cite{bengtsson1999optimal, bengtsson2001optimal}. However, if the number of additional shaping constraints is large, the relaxation is not guaranteed to be tight and a general rank SDR solution may be obtained resulting in suboptimal or even infeasible beamforming solutions. It is demonstrated in \cite{huang2010rank, huang2010dual} that when the number of additional shaping constraints is upper bounded by two, a rank-one solution can always be found by applying a particular rank reduction algorithm, without losing the optimality of the solution. In other cases, if the solution after rank reduction still exceeds the rank of one, a suboptimal solution can be generated from the SDR solution by using, e.g., randomization techniques \cite{karipidis2008quality,luo2010semidefinite}.

In this work, we develop a new approach to optimally solve the downlink beamforming problem in the case that the number of additional shaping constraints is no greater than 79. We exploit CSI knowledge at the transmitter and combine downlink beamforming with full-rate high dimensional real-valued OSTBC to increase the degrees of freedom in the beamformer design. Several works have proposed the idea of combining beamforming with space time coding \cite{George, Zhou1, Zhou2, Li, wu2012rank, wu2013rank, wen2012rank, schad2012convex, Law, Jii}.

In \cite{George}, side information in the form of channel estimates is utilized to design linear beamformers for OSTBC precoded transmission based on a pairwise error probability (PEP) criterion. Two-directional Eigen-beamforming based on channel mean feedback is investigated in \cite{Zhou1} using beamforming along with Alamouti coding \cite{Alamouti1998}, and the symbol error probability (SER) criterion is employed in the beamformer design. A similar idea is applied in \cite{Zhou2} where based on the SER criterion an Eigen-beamformer is designed that exploits the knowledge of channel correlation available at the transmitter. The authors in \cite{Li} consider the same problem as in \cite{George}, yet quasi-orthogonal space time block coding (QSTBC) based beamforming is used instead of OSTBC.

All works of \cite{George, Zhou1, Zhou2, Li} consider the single-user multiple-input and multiple-output (MIMO) scenario. Recently, rank-two beamforming approaches have been independently proposed in \cite{wu2012rank, wu2013rank} and \cite{wen2012rank} to enhance conventional single-group multicast beamforming in which multiple users are served on the same frequency resource subject to transmitted power and QoS constraints. A similar approach of rank-two beamforming is proposed for single-group multicasting using a relay network in \cite{schad2012convex}. The concept of rank-two beamforming
has later been extended in \cite{Law} and \cite{Jii} to solve the multi-group multicast beamforming problem in which beamforming is used to simultaneously deliver independent data services to users in predefined subscriber groups.

By combining Alamouti coding with beamforming, rank-two beamforming approaches outperform the conventional rank-one approaches, however, the drawback associated with these rank-two beamforming approaches is that an optimal solution can only be obtained if the SDR solution\footnote{With SDR solution we refer to the output obtained from interior point solvers such as CVX \cite{grant2008cvx,boyd2004convex}.} exhibits a rank less than or equal to two, otherwise, an approximate solution is obtained in general. As discussed in \cite{wu2013rank}, when the rank of the SDR solution is greater than two, high dimensional ($>\!\! 2$) complex OSTBC can be applied instead of Alamouti coding to preserve the optimality of the beamforming solution, however at the expense of a reduced symbol rate associated with these OSTBC schemes \cite{tarokh1999space}.

The idea of combining beamforming with OSTBC in this work follows the general framework of \cite{wen2012rank, wu2012rank, wu2013rank, schad2012convex, Law, Jii} in which rank-two beamformers are designed in combination with the application of Alamouti coding \cite{Alamouti1998}. In contrast to the rank-two beamformer designs in \cite{wen2012rank, wu2012rank, wu2013rank, schad2012convex, Law, Jii}, we consider herein the downlink beamforming problem where each user is designed to be served by multiple beamformers combined with full-rate higher dimensional real-valued OSTBC. Real-valued OSTBC is employed because of its full (symbol)-rate property, thus the general rank approach proposed in the work achieves full-rate transmission as in the rank-one approaches of \cite{huang2010rank, huang2010dual} and rank-two approaches of \cite{wen2012rank, wu2012rank, wu2013rank, schad2012convex, Law, Jii}.

In order to combine downlink beamforming with real-valued OSTBC, the effective channel vector of each user is adjusted to result in a real vector by applying a phase rotation procedure to which the optimal beamforming solution is proven to be invariant. Due to the orthogonality of the real-valued OSTBC, symbol by symbol detection can be performed at the receivers and the decoding complexity is not increased as compared to the conventional transmission that does not employ OSTBC. The use of OSTBC results in multiple beamformers at each user and therefore multiplies the degrees of freedom in the beamformer design offering improved beamforming performance. Interestingly, the proposed beamformer design disembogues in the same SDR formulation as obtained in the conventional rank-one beamforming approaches of \cite{huang2010rank, huang2010dual} and the rank-two beamforming approaches of \cite{wen2012rank, wu2012rank, wu2013rank, schad2012convex, Law, Jii}, i.e., the beamforming problems after rank relaxation become identical.

In the case that in the conventional rank-one downlink beamforming problem the rank of the SDR solution is greater than one, a rank reduction technique is applied to reduce the rank \cite{huang2010rank, huang2010dual, Pataki}. Similarly, in the Alamouti coding based beamforming approaches of \cite{wen2012rank, wu2012rank, wu2013rank, schad2012convex, Law, Jii}, rank reduction is applied if the SDR solution exhibits a rank greater than two. In our proposed real-valued OSTBC based beamforming approach, the SDR solution after the rank reduction procedure is proven to be optimal for the original problem if all ranks are no greater than eight. In the case that the SDR solution after rank reduction exhibits a rank greater than eight, randomization techniques can be applied to compute an approximate solution \cite{karipidis2008quality,luo2010semidefinite}. Furthermore, we analytically prove that in our approach an optimal solution is always attainable if the number of additional shaping constraints does not exceed 79, whereas in the conventional rank-one approach in \cite{huang2010rank, huang2010dual} and rank-two approach in \cite{wu2012rank, wen2012rank, wu2013rank,schad2012convex, Law, Jii}, the maximal numbers of the shaping constraints are restricted to two and seven, respectively. Simulation results demonstrate the advantage of the proposed approach.

The contribution of this paper can be summarized as follows:
\begin{itemize}
  \item We address the problem of optimal QoS based downlink beamforming in the presence of a massive number of arbitrary quadratic shaping constraints.
  \item We increase the degrees of freedom in the beamformer design by combining optimal linear downlink beamforming with high dimensional real-valued OSTBC exploiting CSI knowledge at the transmitter. Our design can be considered as a nontrivial full-rate extension of the Alamouti coding based rank-two beamforming framework of \cite{wu2012rank, wu2013rank, wen2012rank, schad2012convex, Law, Jii} to general rank beamforming supporting up to eight beamformers per user.
  \item We analytically prove that in our approach an optimal beamforming solution can always be obtained when the number of additional shaping constraints does not exceed $79$.
  \item Extensive simulation results demonstrate the effectiveness of the proposed downlink beamforming scheme in scenarios where the numbers of additional shaping constraints are extremely large.
\end{itemize}

The remainder of the paper is organized as follows. Section II introduces the signal model and revises the conventional rank-one downlink beamforming problem. In Section III, the system model corresponding to the real-valued OSTBC based general rank beamforming approach is developed. Section IV formulates the optimal downlink beamforming problem involving real-valued OSTBC and provides the SDR solution. Section V addresses the problem of computing optimal beamforming vectors from the SDR solution and provides a theoretic analysis regarding the optimality of the proposed downlink beamforming design. Simulation results are carried out in Section VI and conclusions are drawn in Section VII.

\textit{Notation:}
${\rm E}(\cdot)$, $\|\cdot\|_2$, $(\cdot)^{T}$, $(\cdot)^{H}$, $\text{rank}(\cdot)$, ${\rm Tr}(\cdot)$, $\text{diag}\{\cdots\}$, $[\cdot]_{ij}$, and $\angle(\cdot)$ denote statistical expectation, the Euclidean norm, the transpose, the Hermitian transpose, the rank of a matrix, the trace of a matrix, the diagonal matrix formed from the elements in the argument, the entry in the $i$-th row and $j$-th column of the matrix in the argument, and the argument of a complex number, respectively. $\mathbf{I}_K$ denotes the $K\times K$ identity matrix and $\mathbf{X} \succeq \mathbf{0}$ means that the matrix is a positive semidefinite matrix. $\unrhd_l$ denotes a sign in the set $\{\geq,\leq,=\}$.

\section{Rank-one Beamforming}
We consider a cellular communication system where the serving base station equipped with an array of $N$ antennas transmits independent information to $M$ single-antenna receivers. Let $s_i$ denote the information symbol for the $i$-th receiver with zero mean and unit variance. In conventional (rank-one) beamforming approaches of \cite{rashid1998, bengtsson1999optimal, bengtsson2001optimal, schubert2004solution, Shamai2006, gershman2010convex, huang2010rank, huang2010dual}, the transmitter sends a superposition of signals $\{s_i\}_{i=1}^M$ for the different receivers using the respective $N\times 1$ beamforming vectors $\{\textbf{w}_i\}_{i=1}^M$. The received signal at the $i$-th single-antenna receiver is then given by \cite{bengtsson1999optimal}
\begin{equation}
y_i = \underbrace{s_i\mathbf{w}_i^{H}\mathbf{h}_i}_\text{desired signal} + \underbrace{\sum\limits_{m=1, m\neq i}^{M} s_m\mathbf{w}_m^{H}\mathbf{h}_i + n_i}_\text{interference plus noise}
\label{eq:conventional_received_signal}
\end{equation}
where $\mathbf{h}_{i}$ and $n_i$ are the $N\times 1$ channel vector containing the flat fading channel conditions and the receiver noise of variance $\sigma_i^2$, respectively. The total transmitted power at the base station equals $\sum \limits_{i=1}^{M} \mathbf{w}_i^H\mathbf{w}_i$. Based on \eqref{eq:conventional_received_signal}, the SINR at the $i$-th receiver is derived as
\begin{align}
	{\rm SINR}_i \triangleq \frac{|\mathbf{w}_i^{H}\mathbf{h}_i|^2}{\sum\limits_{m=1, m\neq i}^{M} |\mathbf{w}_m^{H}\mathbf{h}_i|^2 + \sigma_i^2}.
\end{align}
Considering a QoS based beamforming design, we define $\gamma_i$ as the minimum SINR requirement of the $i$-th user. Then the extended downlink beamforming problem of minimizing the total transmitted power subject to minimum SINR constraints for each user and additional context specific shaping constraints can be formulated as \cite{huang2010rank, huang2010dual}
\begin{subequations}
\label{eq:conventional_problem}
\begin{eqnarray}
\min_{{\{\mathbf{w}_i\}_{i=1}^{M}}}\!\!\!&&\!\!\!\!\!\! \sum \limits_{i=1}^{M} \mathbf{w}_i^H\mathbf{w}_i \\
\text{s.t.} \!\!\!&&\!\!\!\!\!\! \sum \limits_{m=1}^{M} \mathbf{w}_m^{H}\mathbf{A}_{im}\mathbf{w}_m \unrhd_i b_i, \forall i=1,\dots,M \label{5b} \\
\!\!\!&& \!\!\!\!\!\! \sum \limits_{m=1}^{M} \mathbf{w}_m^{H}\mathbf{A}_{lm}\mathbf{w}_m \unrhd_l b_l, \forall l\!=\!\!M\!\!+\!\!1,\dots,M\!\!+\!\!L \label{5c}
\end{eqnarray}
\end{subequations}
where \eqref{5b} represents a well-known reformulation of the QoS constraints with
\begin{align}
\mathbf{A}_{im}  &\triangleq \begin{cases} \mathbf{h}_i\mathbf{h}_i^H & \text{$m=i,\quad \forall i,m=1,\dots,M$}
\\
-\gamma_i\mathbf{h}_i\mathbf{h}_i^H &\text{$m\not = i,\quad \forall i,m=1,\dots,M$}
\label{eq:Amatrix}
\end{cases}
\end{align}
and
\begin{align}
b_i  &\triangleq \gamma_i\sigma_i^2,~ \unrhd_i  \triangleq \;\geq, ~~\forall i=1,\dots,M
\label{eq:bisymbol}
\end{align}
and $L$ additional quadratic shaping constraints are formulated in \eqref{5c} for appropriately chosen (as specified below) $N \times N$ Hermitian matrices
$\mathbf{A}_{lm}, \forall  l = M\!+\!1,\ldots,M\!+\!L; \forall m = 1,\ldots, M$, that are not necessarily positive definite, with corresponding thresholds
$b_l , \forall l = M +1, ...,M +L$. Note that the shaping constraints in \eqref{5c} are not for information decoding purpose, while the first $M$ constraints in \eqref{5b} are.
Depending on the specific application under consideration, the additional shaping constraints in \eqref{5c} may take different forms (c.f. \cite{huang2010rank, huang2010dual}). Popular example applications which can be formulated under the framework of problem (\ref {eq:conventional_problem}) are described in subsections II.A and II.B.

\subsection{Positive Semidefinite Shaping Constraints}
In the context of cognitive radio networks, $\mathbf{A}_{lm} \triangleq \mathbf{h}_l\mathbf{h}^H_l$, where $\mathbf{h}_l$ denotes the channel vector between the base station and the $l$-th primary user. In this case, with $b_l$ denoting the upper power threshold at the primary user and choosing $\unrhd_l\triangleq \; \leq$,  the $l$-th general shaping constraint \eqref{5c} takes the form \footnote{Note that if all matrices $\mathbf{A}_{lm}$ in (\ref{5c}) are positive semidefinite, then the problem (\ref{eq:conventional_problem}) can be reformulated as a SOCP problem.}
\begin{align}
	\sum \limits_{m=1}^{M} \mathbf{w}_m^{H}\mathbf{A}_{lm}\mathbf{w}_m \leq b_l.
	\label{shaping1}
\end{align}
Thus the interference constraint \eqref{shaping1} is used to guarantee that the power leakage to the primary users is below certain threshold \cite{cognitiveradio1, cognitiveradio2, cognitiveradio3}. In the context of femtocell networks, $\mathbf{h}_l$ denotes the channel vector between the base station and the $l$-th concurrent user, and the shaping constraint \eqref{shaping1} is designed to ensure that the power leakage to concurrent users in coexisting hierarchical networks is below certain threshold \cite{femtocell1, femtocell2, femtocell3}.

In the context of physical layer secrecy networks, in contrast, $\mathbf{h}_l$ denotes the channel vector between the base station and the $l$-th eavesdropper, and the shaping constraint \eqref{shaping1} is employed to enforce that the power leakage to eavesdroppers is below certain threshold \cite{physicalsecurity1, physicalsecurity2, physicalsecurity3}.

Similarly, in the context of energy harvesting networks, $\mathbf{h}_l$ denotes the channel vector between the base station and the $l$-th charging terminal \cite{wirelesscharging1, wirelesscharging2, wirelesscharging3}. In this case $b_l$ denotes the minimum power threshold to be guaranteed at the charging terminal and $\unrhd_l\triangleq \; \geq$ is chosen. The shaping constraint \eqref{5c} can be rewritten as
\begin{align}
\sum \limits_{m=1}^{M} \mathbf{w}_m^{H}\mathbf{A}_{lm}\mathbf{w}_m \geq b_l
\label{shaping2}
\end{align}
with $\mathbf{A}_{lm} \triangleq \mathbf{h}_l  \mathbf{h}_l^H$ for $m=1,...,M$ and $l=M+1,...,M+L$ to ensure efficient wireless charging.

\subsection{Indefinite Shaping Constraints}
Indefinite shaping constraints can be used to perform relaxed nulling, as proposed in \cite{nphard}, to reduce intercell interference in multiuser downlink networks. Let $\mathbf{h}_l$ denote the channel vector from the base station of a given serving cell to a user of a different cell for which the interference shall be limited. Defining $\mathbf{A}_{lm} \triangleq \beta\textbf{I}_N-\frac{\textbf{h}_l\textbf{h}_l^H}{\left\|\textbf{h}_l\right\|^2}$, $b_l=0$, and choosing $\beta$ as an appropriate interference threshold parameter \cite{hammarwall2006}, the shaping constraint \eqref{5c} takes the form
\begin{align}
\mathbf{w}_m^{H}\mathbf{A}_{lm}\mathbf{w}_m \geq b_l.
\end{align}
In this design the tolerable interference power induced by the $l$-th user to the $m$-th user depends on the spatial signature $\mathbf{h}_l$ of the co-channel user. Some other applications of indefinite shaping constraints can be found in \cite{hammarwall2006, hammarwall2005}.

\subsection{Semidefinite Relaxation}
In this subsection we briefly revisit the SDR approach that is widely used to approximately solve the beamforming problem of form (\ref {eq:conventional_problem}). The power minimization problem \eqref{eq:conventional_problem} is a quadratically constrained quadratic programming (QCQP) problem which is NP-hard in general \cite{luo2010semidefinite}. Denote $\mathbf{X}_i\triangleq \mathbf{w}_i\mathbf{w}_i^H$, problem (\ref{eq:conventional_problem}) can be rewritten as
\begin{eqnarray}
\min_{{\{\mathbf{X}_i\}_{i=1}^{M}}} \!\!\!\!&&\!\!\!\! \sum \limits_{i=1}^{M} {\rm Tr}(\mathbf{X}_i) \nonumber \\
\text{s.t.}
\!\!\!\!&&\!\!\!\! \sum \limits_{m=1}^{M} {\rm Tr}(\mathbf{A}_{lm}\mathbf{X}_m) \unrhd_l b_l, \forall l =1,\dots,M+L\nonumber \\
\!\!\!\!&&\!\!\!\! \mathbf{X}_i \succeq \mathbf{0}, {\rm rank}(\mathbf{X}_i) = 1, \forall i =1,\dots,M.
\label{eq:conventional_problem2}
\end{eqnarray}
The SDR technique can be employed to solve the convex relaxation of problem \eqref{eq:conventional_problem2} by removing the rank constraints \cite{bengtsson1999optimal, bengtsson2001optimal}. Since the SDR solution is not of rank one in general, rank reduction techniques are applied to obtain a solution to problem \eqref{eq:conventional_problem2} with a reduced rank \cite{huang2010rank,huang2010dual}, see also Section V.A. However, in the case that a rank-one solution does not exist, an approximate solution can be computed from the SDR solution using, e.g., the popular randomization procedures of \cite{karipidis2008quality} and \cite{luo2010semidefinite}.

\section{General Rank Beamforming}
The central idea of combining optimal downlink beamforming with the concept of real-valued OSTBC proposed in this work follows the general framework of \cite{wen2012rank, wu2012rank, wu2013rank, schad2012convex, Law, Jii} in which rank-two beamformers are designed by combining beamforming with Alamouti coding and making  use of CSI available at the transmitter. As compared to the rank-two approaches, we employ full-rate real-valued OSTBC to further increase the degrees of freedom in the beamformer design which grow linearly with the size of the code. Extending the rank-two beamforming approach to higher dimensional ($>\!\!2$) OSTBC has previously been considered as impractical due to the rate penalty associated with these codes \cite{wu2013rank}. By applying real-valued OSTBC at the transmitter, multiple beamformers can be used to deliver the data stream to each user while maintaining the full-rate transmission property.

\subsection{Full-rate Real-valued OSTBC}
Let $\mathcal{X}(\mathbf{u})$ be a $K\times K$ real-valued OSTBC matrix given by \cite{tarokh1999space}
\begin{equation}
 \mathcal{X}(\mathbf{u}) = \sum \limits_{k=1}^{K} u_k \mathbf{C}_k
 \label{eq:realOSTBC}
\end{equation}
where $K$ is the number of symbols per block, $\mathbf{u}\triangleq[u_1,\dots,u_K]^{T}$ is an arbitrary $K\times 1$ real vector and $\mathbf{C}_k$ is a $K\times K$ real code coefficient matrix. Per definition the OSTBC matrix $\mathcal{X}(\mathbf{u})$ satisfies the orthogonality property
\begin{equation}
	\mathcal{X}^{H}(\mathbf{\mathbf{u}}) \mathcal{X}(\mathbf{u}) = \mathcal{X}(\mathbf{u})\mathcal{X}^{H}(\mathbf{u})= \|\mathbf{u}\|_2^2\mathbf{I}_{K}
	\label{eq:orthogonal}
\end{equation}
which will be used in the following subsection. In this paper, we only consider real-valued OSTBC matrices with $K\!=1,2, 4$ or $8$ which are the only possible sizes to achieve full rate \cite{tarokh1999space}. We note that, for a complex symbol vector $\mathbf{u}$, the orthogonality property in \eqref{eq:orthogonal} can only be satisfied if $K\leq 2$ \cite{Jafarkhani:2010:SCT:1841144}. Examples for real-valued OSTBC matrices are
\begin{equation}
\mathcal{X}([u_1,u_2]^{T}) \triangleq \begin{bmatrix}u_1&u_2\\-u_2&u_1 \end{bmatrix},
\label{eq:two_real}
\end{equation}
\begin{equation}
\mathcal{X}([u_1,u_2,u_3,u_4]^{T}) \triangleq \begin{bmatrix}
u_1&u_2&u_3&u_4\\
-u_2&u_1&-u_4&u_3\\
-u_3&u_4&u_1&-u_2\\
-u_4&-u_3&u_2&u_1
\end{bmatrix},
\label{eq:four_real}
\end{equation}
and $\mathcal{X}([u_1,\dots,u_8]^{T})\triangleq$
\begin{equation}
 \begin{bmatrix}
u_1&u_2&u_3&u_4&u_5&u_6&u_7&u_8\\
-u_2&u_1&u_4&-u_3&u_6&-u_5&-u_8&u_7\\
-u_3&-u_4&u_1&u_2&u_7&u_8&-u_5&-u_6\\
-u_4&u_3&-u_2&u_1&u_8&-u_7&u_6&-u_5\\
-u_5&-u_6&-u_7&-u_8&u_1&u_2&u_3&u_4\\
-u_6&u_5&-u_8&u_7&-u_2&u_1&-u_4&u_3\\
-u_7&u_8&u_5&-u_6&-u_3&u_4&u_1&-u_2\\
-u_8&-u_7&u_6&u_5&-u_4&-u_3&u_2&u_1
\end{bmatrix}.
\label{eq:eight_real}
\end{equation}

\subsection{System Model}
Denote $\mathbf{s}_i=[s_{i1},\ldots,s_{iK}]^T$ as the $K\times 1$ complex symbol vector for the $i$-th user with $K\leq N$ and $K\in\{1, 2, 4, 8\}$, i.e., in correspondence with the dimension of the real-valued OSTBC matrices in \eqref{eq:two_real}-\eqref{eq:eight_real}.
In this work, we employ the real-valued OSTBC structure $\mathcal{X}(\cdot)$ given in (\ref{eq:realOSTBC}) on the complex symbol vector $\mathbf{s}_i$. Instead of weighting each symbol by a beamforming vector as in (\ref{eq:conventional_received_signal}), a code matrix $\mathcal{X}(\mathbf{s}_i)$ is transmitted for each user applying $K$ beamformers of length $N$, denoted as $\textbf{w}_{i1},\ldots,\textbf{w}_{iK}$. In this case, taking a slightly different perspective, each of the $K$ beams can be regarded as a virtual antenna from which the OSTBC is transmitted. In our scenario we consider a block fading channel model where the channels remain constant over $K$ time slots. The received signal $y_{ik}$ at the $i$-th user in the $k$-th time slot is given by
\begin{eqnarray}
	y_{ik} &=& \sum \limits_{m=1}^{M} \sum \limits_{k'=1}^{K}[\mathcal{X}(\mathbf{s}_m)]_{kk'}\mathbf{w}_{mk'}^{H}\mathbf{h}_{i} + n_{ik}
	\label{eq:receive_signal}
\end{eqnarray}
where $n_{ik}$ is the noise of the $i$-th user in the $k$-th time slot.
In a compact matrix notation, the received signal vector $\mathbf{y}_i$$\triangleq$$[y_{i1},\dots,y_{iK}]^{T}$ at the $i$-th user within the transmission period of $K$ time slots is given by
\begin{eqnarray}
	\mathbf{y}_i &=& \sum \limits_{m=1}^{M}\mathcal{X}(\mathbf{s}_m)\mathbf{W}_m^{H}\mathbf{h}_{i} + \mathbf{n}_i\nonumber \\
	&=& \underbrace{\mathcal{X}(\mathbf{s}_i)\mathbf{W}_i^{H}\mathbf{h}_{i}}_\text{desired signal} + 
	\underbrace{\sum \limits_{m=1, m\neq i}^{M} \mathcal{X}(\mathbf{s}_m) \mathbf{W}_m^{H}\mathbf{h}_{i} + \mathbf{n}_i}_\text{interference plus noise} \nonumber\\
	\label{eq:receive_vector}
\end{eqnarray}
where
\begin{equation}
\textbf{W}_{i} \triangleq [\textbf{w}_{i1},\ldots,\textbf{w}_{iK}], \quad K \in \{1,2,4,8\}
\end{equation}
is the beamforming matrix, and $\mathbf{n}_i$$\triangleq$$ [n_{i1},\dots,n_{iK}]^{T}$. We assume that the noise vector $\mathbf{n}_i$ at the $i$-th receiver is zero mean spatially and temporally white circular complex Gaussian with covariance matrix $\sigma^2_i \mathbf{I}_{K}$. The above system model can be reformulated in the following equivalent form \cite{Jafarkhani:2010:SCT:1841144}
\begin{eqnarray}
	\mathbf{\tilde{y}}_i \!\!\!\!&=\!& \sum \limits_{m=1}^{M}\mathcal{X}(\mathbf{W}_m^{H}\mathbf{h}_{i})\mathbf{s}_i + \mathbf{\tilde{n}}_i\nonumber \\
	&=\!& \mathcal{X}(\mathbf{W}_i^{H}\mathbf{h}_{i})\mathbf{s}_i \!+\! 
	\underbrace{\sum \limits_{m=1, m\neq i}^{M} \mathcal{X}(\mathbf{W}_m^{H}\mathbf{h}_{i})\mathbf{s}_m}_{\mathbf{\tilde{i}}_i} +\mathbf{\tilde{n}}_i
	\label{eq:reformulated_receive}
\end{eqnarray}
where
\begin{eqnarray}
\mathbf{\tilde{y}}_i & \triangleq & \begin{bmatrix}y_{i1},-y_{i2},\dots,-y_{iK}\end{bmatrix}^{T},\\
\mathbf{\tilde{i}}_i & \triangleq & \sum \limits_{m=1, m\neq i}^{M} \mathcal{X}(\mathbf{W}_m^{H}\mathbf{h}_{i})\, \mathbf{s}_m, \\
\mathbf{\tilde{n}}_i & \triangleq & \begin{bmatrix}n_{i1},-n_{i2},\dots,-n_{iK}\end{bmatrix}^{T}.	
\end{eqnarray}
In order to implement full-rate transmission and symbol-wise decoding for each user, the code matrix $\mathcal{X}(\mathbf{W}_i^{H}\mathbf{h}_{i})$ has to exhibit the orthogonality property \eqref{eq:orthogonal}. This however requires that the virtual channel vector $\mathbf{W}_i^{H}\mathbf{h}_{i}$ becomes real-valued, i.e., the condition
 \begin{align}
 \mathbf{W}_i^{H}\mathbf{h}_i \in \mathbb{R}^{K},~~~\quad \forall i=1,\dots,M
 \label{phaseconstraints1}
 \end{align}
 holds. We remark that in general $\mathbf{W}_m^H\mathbf{h}_i$ is not real-valued for $m \neq i$, and thus $\mathcal{X}(\mathbf{W}_m^{H}\mathbf{h}_{i})$ does not necessarily satisfy the orthogonal property in \eqref{eq:orthogonal}. Note that our proposed scheme is associated with a slight increase in the signaling overhead. This is due to the fact that the receivers have to know their individual composite channels for the decoding, i.e., the $i$th user requires the knowledge of $\textbf{W}_i^H\mathbf{h}_i$. This signaling overhead is also required in the rank-one and rank-two schemes, however, the amount of signaling linearly grows with the code dimension of $K$. In this sense, the increase in the signaling overhead for our proposed scheme only applies to cases where there is a clear benefit in using general rank beamforming. In the following, we derive explicit expressions for the post detection SINR of the symbols received at the destinations under the assumption that condition \eqref{phaseconstraints1} is satisfied and that signal user detection is applied at the receivers. Towards this aim, we introduce the following useful lemma.
\newtheorem*{lemma1}{Lemma 1}
\begin{lemma1}
Assume that $\boldsymbol\psi$ and $\boldsymbol\omega$ are a real and a complex vector both with the dimension $K\times 1$, respectively. Let $\mathbf{\Phi}\triangleq\mathcal{X}^H(\boldsymbol\psi)\mathcal{X}(\boldsymbol\omega)\mathcal{X}^{H}(\boldsymbol\omega)\mathcal{X}(\boldsymbol\psi)$ where $\mathcal{X}(\cdot)$ is a $K\times K$ real-valued OSTBC structure that fulfils (\ref{eq:orthogonal}). Then
\begin{equation}
	[\mathbf{\Phi}]_{kk} = \|\boldsymbol\psi\|_2^2 \|\boldsymbol\omega\|_2^2 \quad \forall k=1,\dots,K
	\label{eq:lemma_1}
\end{equation}
where $[\mathbf{\Phi}]_{kk}$ is the $k$-th diagonal element of the matrix $\mathbf{\Phi}$.
\end{lemma1}
\begin{IEEEproof}
Let $\mathcal{X}(\boldsymbol\omega)\triangleq\mathbf{\Omega}_1+j\mathbf{\Omega}_{2}$ where $\mathbf{\Omega}_{1}$ and $\mathbf{\Omega}_{2}$ are real orthogonal matrices from the definition of  $\mathcal{X}(\boldsymbol\omega)$. Then
\begin{eqnarray}\mathcal{X}(\boldsymbol\omega)\mathcal{X}^{H}(\boldsymbol\omega)\!\!&=&\!\!(\mathbf{\Omega}_{1}\mathbf{\Omega}_{1}^{T}+\mathbf{\Omega}_{2}\mathbf{\Omega}_{2}^{T}) + j(\mathbf{\Omega}_{2}\mathbf{\Omega}_{1}^{T}-\mathbf{\Omega}_{1}\mathbf{\Omega}_{2}^{T}) \nonumber\\
\!\!&=& \!\!\|\boldsymbol\omega\|_2^2\mathbf{I}_R+j(\mathbf{\Omega}_{2}\mathbf{\Omega}_{1}^{T}-\mathbf{\Omega}_{1}\mathbf{\Omega}_{2}^{T}).
\end{eqnarray}
Hence
\begin{eqnarray}
\mathbf{\Phi}\!\!\! &=& \!\!\! \|\boldsymbol\omega\|_2^2\mathcal{X}^H(\boldsymbol\psi)\mathcal{X}(\boldsymbol\psi) \!+ \!j\mathcal{X}^H(\boldsymbol\psi)(\mathbf{\Omega}_{2}\mathbf{\Omega}_{1}^{T}-\mathbf{\Omega}_{1}\mathbf{\Omega}_{2}^{T})\mathcal{X}(\boldsymbol\psi) \nonumber\\
\!\!\!&=& \!\!\!\|\boldsymbol\psi\|_2^2\|\boldsymbol\omega\|_2^2\mathbf{I}_K \!+\! j\mathcal{X}^H(\boldsymbol\psi)(\mathbf{\Omega}_{2}\mathbf{\Omega}_{1}^{T}-\mathbf{\Omega}_{1}\mathbf{\Omega}_{2}^{T})\mathcal{X}(\boldsymbol\psi).
\end{eqnarray}
Since $\mathbf{\Phi}$ is a Hermitian matrix and $\mathcal{X}(\boldsymbol\psi)$ is a real matrix, $\mathcal{X}^H(\boldsymbol\psi)(\mathbf{\Omega}_{2}\mathbf{\Omega}_{1}^{T}-\mathbf{\Omega}_{1}\mathbf{\Omega}_{2}^{T})\mathcal{X}(\boldsymbol\psi)$ is a skew symmetric matrix, i.e., its elements on the main diagonal are zero. Then the equation (\ref{eq:lemma_1}) holds.
\end{IEEEproof}
For an orthogonal matrix $\mathcal X(\mathbf{W}_i^{H}\mathbf{h}_{i})$, i.e., with $\mathbf{W}_i^{H}\mathbf{h}_{i}$ satisfying \eqref{phaseconstraints1}, the transmitted symbol vector can be equalized as
 \begin{eqnarray}
 	\mathbf{\hat{s}}_{i} &=& \frac{1}{\|\mathbf{W}_i^{H}\mathbf{h}_{i}\|_2^2}\mathcal{X}^{H}(\mathbf{W}_i^{H}\mathbf{h}_{i}) \mathbf{\tilde{{y}}}_i\nonumber\\
 	 &=&\mathbf{s}_i + \frac{1}{\|\mathbf{W}_i^{H}\mathbf{h}_{i}\|_2^2}\mathcal{X}^{H}(\mathbf{W}_i^{H}\mathbf{h}_{i})(\mathbf{\tilde{i}}_i + \mathbf{\tilde{n}}_i).
 \label{eq:matched}
 \end{eqnarray}
 Based on \eqref{eq:matched}, the covariance matrix of the received interference contained in $\mathbf{\hat{s}}_{i}$ is given by
 \begin{eqnarray}
 	\mathbf{R}_{i}^{(\textup{I})}&=& \frac{1}{\|\mathbf{W}_i^{H}\mathbf{h}_{i}\|_2^4}\mathcal{X}^{H}(\mathbf{W}_i^{H}\mathbf{h}_{i}){\rm E}\{\mathbf{\tilde{i}}_i\mathbf{\tilde{i}}_i^{H}\}\mathcal{X}(\mathbf{W}_i^{H}\mathbf{h}_{i})\nonumber\\
	&=& \frac{1}{\|\mathbf{W}_i^{H}\mathbf{h}_{i}\|_2^4}[\sum \limits_{m=1, m\neq i}^{M}\mathcal{X}^{H}(\mathbf{W}_i^{H}\mathbf{h}_{i})\mathcal{X}(\mathbf{W}_m^{H}\mathbf{h}_{i})\times \nonumber\\
 	&&\mathcal{X}^{H}(\mathbf{W}_m^{H}\mathbf{h}_{i})\mathcal{X}(\mathbf{W}_i^{H}\mathbf{h}_{i})]
 	\label{eq:new_covariance}
 \end{eqnarray}
 and the covariance matrix of the noise in $\mathbf{\hat{s}}_{i}$ is given by
 \begin{eqnarray}
 	\mathbf{R}_{i}^{(\textup{N})}&=& \frac{1}{\|\mathbf{W}_i^{H}\mathbf{h}_{i}\|_2^4}\mathcal{X}^{H}(\mathbf{W}_i^{H}\mathbf{h}_{i}){\rm E}\{\mathbf{\tilde{n}}_i\mathbf{\tilde{n}}_i^{H}\}\mathcal{X}(\mathbf{W}_i^{H}\mathbf{h}_{i})\nonumber\\
	&=&\frac{\sigma_i^2}{\|\mathbf{W}_i^{H}\mathbf{h}_{i}\|_2^2}\mathbf{I}_{K}.
 	\label{eq:noise}
 \end{eqnarray}
Substituting according to (\ref {phaseconstraints1}) the real-valued vector $\boldsymbol\psi=\mathbf{W}_i^{H}\mathbf{h}_{i}$ and complex vector $\boldsymbol\omega=\mathbf{W}_m^{H}\mathbf{h}_{i}$
in (\ref{eq:new_covariance}) and applying Lemma $1$, the interference power of the $i$-th user in the $k$-th time slot can be expressed as
\begin{eqnarray}
[\mathbf{R}_{i}^{(\textup{I})}]_{kk} &=& \frac{1}{\|\mathbf{W}_i^{H}\mathbf{h}_{i}\|_2^2}\sum \limits_{m\neq i} \|\mathbf{W}_m^{H}\mathbf{h}_{i}\|_2^2.
\label{eq:k_th_interference_matrix}
\end{eqnarray}
With \eqref{eq:noise} and \eqref{eq:k_th_interference_matrix}, the
post detection SINR corresponding to symbol $s_{ik}$ is given by
\begin{eqnarray}
	\text{SINR}(s_{ik}) & \triangleq & \frac{{\rm E}\{s_{ik}s_{ik}^{\star} \}}{[\mathbf{R}_{i}^{(\textup{I})}]_{kk} + [\mathbf{R}_{i}^{(\textup{N})}]_{kk}}\nonumber\\
	&=& \frac{\|\mathbf{W}_i^{H}\mathbf{h}_{i}\|_2^2}{\sum \limits_{m=1, m\neq i}^{M} \|\mathbf{W}_m^{H}\mathbf{h}_{i}\|_2^2 + \sigma_i^2}.
\label{eq:sinr}
\end{eqnarray}
We note that the expression in \eqref{eq:sinr} is independent of the time index $k$ and hence for the $i$-th user the post detection SINR is identical for all symbols in the OSTBC block. Thus we denote $\text{SINR}_i=\text{SINR}(s_{ik})$. For simplicity of presentation, the SINR constraints in the general rank approach can be written in a similar form as in the rank-one beamforming approach of \eqref{5b}, i.e.,
\begin{equation}
\sum \limits_{m=1}^{M} {\rm Tr}(\mathbf{A}_{im}\mathbf{W}_m\mathbf{W}_m^{H}) \unrhd_i b_i \quad \forall i = 1,\dots,M
\label{eq:sinr_constraints}
\end{equation}
where $\mathbf{A}_{im}$ is defined in (\ref{eq:Amatrix}).

Since each symbol appears only once in each row of the code matrix $\mathcal{X}(\mathbf{s}_i)$, c.f. \eqref{eq:two_real}-\eqref{eq:eight_real}, the transmitted power towards the $i$-th user in the $k$-th time slot can be computed as
\begin{eqnarray}
	P_{ik} &=& {\rm E}\{\mathbf{e}_k^{H}\mathcal{X}(\mathbf{s}_i) \mathbf{W}_i^{H}\mathbf{W}_i\mathcal{X}^{H}(\mathbf{s}_i)\mathbf{e}_k\}\nonumber\\
	&=& {\rm Tr}(\mathbf{W}_i{\rm E}\{\mathcal{X}^{H}(\mathbf{s}_i)\mathbf{e}_k\mathbf{e}_k^{H}\mathcal{X}(\mathbf{s}_i)\}\mathbf{W}_i^{H})\nonumber\\
	&=&{\rm Tr}(\mathbf{W}_i\mathbf{W}_i^{H})
\label{eq:transmit_power_i}
\end{eqnarray}
where $\mathbf{e}_k$ is the $k$-th column of the $N \times N$ identity matrix. Similarly we observe that the transmitted power $P_{ik}$ is identical in all $K$ time slots. Let $P_i = P_{ik}$ represent the transmitted power towards the $i$-th user in each time slot. Then the total transmitted power in each time slot amounts to
\begin{equation}
\sum \limits_{i=1}^{M} P_i = \sum \limits_{i=1}^{M} {\rm Tr}(\mathbf{W}_i\mathbf{W}_i^{H}).
\label{eq:total_transmit_power}
\end{equation}
With multiple beamformers designed for each user instead of a single one, the additional shaping constraints in (\ref{5c}) can be expressed as
\begin{eqnarray}
 &&\sum_{m=1}^M \sum_{k=1}^K \textbf{w}_{mk}^H \textbf{A}_{lm} \textbf{w}_{mk}  \nonumber \\
 =&&\sum_{m=1}^M {\rm Tr}(\textbf{W}_m^H\textbf{A}_{lm}\textbf{W}_m) \nonumber \\
=&& \sum \limits_{m=1}^{M} {\rm Tr}(\mathbf{A}_{lm}\mathbf{W}_m\mathbf{W}_m^{H}) \unrhd_l b_l \nonumber\\
&&\forall l=M+1,\dots,M+L.
\label{eq:general_matrix}
\end{eqnarray}
\section{The Power Minimization Problem}
In this section, we consider the problem of minimizing the total transmitted power per time slot subject to SINR constraints at each user and additional shaping constraints on the beamformers. Taking into account that according to \eqref{phaseconstraints1} the virtual channel vector $\textbf{W}_i^{H}\textbf{h}_i$ must be real-valued in order to satisfy the orthogonality property for simple decoding, the optimization problem is formulated in the following form
\begin{subequations}
\label{eq:original1}
\begin{eqnarray}
	\min_{\{\mathbf{W}_i\}_{i=1}^{M}} \!\!&& \!\! \sum \limits_{i=1}^{M} {\rm Tr}(\mathbf{W}_i\mathbf{W}_i^{H})\\
	\text{s.t.} \!\!&& \!\! \sum \limits_{m=1}^{M} {\rm Tr}(\mathbf{A}_{lm}\mathbf{W}_m\mathbf{W}_m^{H}) \unrhd_l b_l \nonumber\\
	&&\quad \forall l=1,\dots,M+L \label{sub:constraints}\\
	 \!\!&& \!\! \mathbf{W}_i^{H}\mathbf{h}_i \in \mathbb{R}^{K}, \quad \forall i=1,\dots,M. \label{sub:real}
\end{eqnarray}
\end{subequations}
We remark that as a special case, the Alamouti code can be employed in our proposed scheme without the need of imposing the constraint \eqref{sub:real}, since the Alamouti code satisfies the orthogonality property of \eqref{eq:orthogonal} for an arbitrary complex vectors $\mathbf{u}$ while achieving full rate. In this case the proposed scheme becomes similar to the rank-two schemes proposed in \cite{wen2012rank, wu2012rank, wu2013rank, schad2012convex, Law, Jii}.

\subsection{Phase Rotation Invariance Property}
To solve the problem \eqref{eq:original1}, we first consider a relaxed problem of \eqref{eq:original1} by removing the constraints \eqref{sub:real}.
Let $\mathbf{W}_i^{\star} \triangleq \begin{bmatrix}\mathbf{w}_1^{i\star},\dots,\mathbf{w}_K^{i\star}\end{bmatrix}$ denote an optimal solution of the relaxed optimization problem of \eqref{eq:original1} in which constraint \eqref{sub:real} is omitted. Then we can perform the phase rotation on $\{\mathbf{W}_i^{\star}\}_{i=1}^M$ according to
\begin{align}
\label{rotation}
\mathbf{W}^{\prime\star}_i \triangleq \mathbf{W}_i^{\star}\boldsymbol{\Theta}_i	
\end{align}
where the diagonal matrix $\boldsymbol{\Theta}_i$ is given by
\begin{align}
	\boldsymbol{\Theta}_i \triangleq \text{diag}\{&\exp(j\angle(\mathbf{w}_{i1}^{\star H}\mathbf{h}_i)),\!\dots\!,\!\exp(j\angle(\mathbf{w}_{iK}^{\star H}\mathbf{h}_i))\}.
\end{align}
Since $\{\mathbf{W}^{\prime\star}_i\}_{i=1}^M$ satisfies all the constraints in \eqref{eq:original1}, including constraint \eqref{sub:real}, it is a feasible solution to the unrelaxed problem \eqref{eq:original1}. As the total transmitted power associated with $\{\mathbf{W}^{\prime\star}_i\}_{i=1}^M$ is the same as that associated with the optimal solution $\{\mathbf{W}_i^{\star}\}_{i=1}^M$, we conclude that $\{\mathbf{W}^{\prime\star}_i\}_{i=1}^M$ is an optimal solution to the original problem \eqref{eq:original1}. In other words, relaxing the real-valued requirements expressed in constraints \eqref{sub:real} in the beamforming problem \eqref{eq:original1} results in an equivalent problem. An optimal solution for the original problem can always be computed from the solution of the relaxed problem by applying the phase rotation proposed in \eqref{rotation}. Therefore, we can, without loss of generality, omit constraint \eqref{sub:real} when solving \eqref{eq:original1}.

\subsection{SDR Approach}
Let us define the variable transformation
\begin{equation}\label{def:Xi}
\mathbf{X}_i \triangleq \mathbf{W}_i\mathbf{W}_i^{H}.
\end{equation}
Then substituting $\mathbf{X}_i$ in \eqref{eq:original1} and adding the constraints
\begin{subequations}
\begin{eqnarray}
  & & \mathbf{X}_i \succeq \mathbf{0} \\
  & & \text{rank}(\mathbf{X}_i) \leq K, \quad \forall i=1,\dots,M
	\label{eq:rank_constraint0}
\end{eqnarray}
\end{subequations}
to ensure that the transformation (\ref{def:Xi}) exists,
problem \eqref{eq:original1} (with relaxed constraint \eqref{sub:real}) converts to a rank constrained semidefinite program
\begin{subequations}
\begin{eqnarray}
\min_{\{\mathbf{X}_i\}_{i=1}^{M}}  \!\!&& \!\!\!\! \sum \limits_{i=1}^{M} {\rm Tr}(\mathbf{X}_i)\\
 \text{s.t.} \!\!&&\!\!\!\! \sum \limits_{m=1}^{M}\! {\rm Tr}(\! \mathbf{A}_{lm}\mathbf{X}_m\!) \!\unrhd_l\! b_l, \! \forall l\!=\!1,\!\dots,\!M\!+\!L  \\
 \!\!&& \!\!\!\! \mathbf{X}_i \succeq \mathbf{0}, \\
 \!\!&& \!\!\!\! \text{rank}(\mathbf{X}_i) \leq K, \quad \forall i=1,\dots,M. \label{eq:rank_constraint}
\end{eqnarray}
\label{eq:rankconstrained_sdp}
\end{subequations}
We remark that problem (\ref{eq:rankconstrained_sdp}) is identical to problem (\ref{eq:conventional_problem2}) except for the rank constraint. While in the latter problem the optimization variable $\mathbf{X}_i$ is restricted to the set of rank-one matrices, in our proposed formulation (\ref{eq:rankconstrained_sdp}) the rank of the matrix must not exceed $K$. This shows that the feasible set of our proposed beamforming approach is greater than that of the conventional one.

Since the rank constraints in \eqref{eq:rankconstrained_sdp} are non-convex, we employ the SDR approach \cite{bengtsson1999optimal} to obtain a relaxed convex optimization problem in which the rank constraints in (\ref{eq:rank_constraint}) are omitted,
\begin{subequations}
\begin{eqnarray}
\min_{\{\mathbf{X}_i\}_{i=1}^{M}}\!\!  &&\!\!  \sum \limits_{i=1}^{M} {\rm Tr}(\mathbf{X}_i)\\
 \text{s.t.} \!\!&&\!\!\! \sum \limits_{m=1}^{M}\! {\rm Tr}(\! \mathbf{A}_{lm}\mathbf{X}_m\!) \!\unrhd_l\! b_l, \! \forall l\!=\!1,\!\dots,\!M\!+\!L \label{eq:mainconstraint}\\
&& \!\mathbf{X}_i \succeq \mathbf{0}, \quad \forall i=1,\dots,M.
\end{eqnarray}
\label{eq:relaxed}
\end{subequations}
For later reference, we also provide the Lagrange dual problem of \eqref{eq:relaxed} which has the following form \cite{huang2010rank}
\begin{eqnarray}
\max_{\{\eta_l\}_{l=1}^{M+L}}&& \sum \limits_{l=1}^{M+L}\eta_l b_l\nonumber\\
\text{s.t.}&& \mathbf{Z}_i = \mathbf{I} - \sum \limits_{l=1}^{M+L}\eta_l\mathbf{A}_{li} \succeq \mathbf{0} \quad \forall i= 1,\dots,M\nonumber\\
&& \eta_l \unrhd_l^{*} 0 \quad \forall l=1,\dots,M+L
\label{eq:dual_problem}
\end{eqnarray}
where
\begin{equation}
\unrhd_l^{*} \triangleq \begin{dcases*} \geq &if $\unrhd_l$ is  $\geq$ \\
\text{unrestricted\footnotemark} &if $\unrhd_l$ is $=$\\
\leq &if $\unrhd_l$ is $\leq$\end{dcases*}.
\end{equation}
\footnotetext{i.e., the constraint is omitted.}
Note that, according to our previous observation, problem \eqref{eq:relaxed} is identical to the SDR of the rank-one beamforming problem \eqref{eq:conventional_problem2}. Therefore, the complexity at the transmitter side does not differ from that of the rank-one and rank-two schemes. This is due to the observation that the computational complexity of the proposed general rank approach mainly consists in solving (\ref{eq:rankconstrained_sdp}), which is the same as in the rank-one and rank-two approaches. Problem \eqref{eq:relaxed} belongs to the class of separable semidefinite programming (SDP) problems \cite{huang2010rank,huang2010dual} and can be solved efficiently using solvers such as CVX \cite{grant2008cvx,boyd2004convex}. Denote $\{\mathbf{X}^{\star}_i\}_{i=1}^{M}$ as an optimal solution to the problem \eqref{eq:relaxed}.
Then we can apply the rank reduction algorithm proposed in \cite{huang2010rank} and \cite{huang2010dual} with the input $\{\mathbf{X}^{\star}_i\}_{i=1}^{M}$ to reduce the rank of the optimal solution. A detailed description of the rank reduction procedure for general rank matrices is provided in Section V.

\section{Beamforming matrices generation}
In this section, we derive a sufficient condition on the maximum number of shaping constraints under which a solution to (35) can always be obtained from the SDR solution. In this context, we adapt the rank reduction algorithm of \cite{huang2010rank}, \cite{huang2010dual} with modified stopping criteria for its application in general rank beamforming. Then we address the issue of determining the smallest code dimension $K$ for all downlink users based on the output of the rank reduction procedure. In the case that a SDR solution after the rank reduction procedure has a rank greater than eight, a randomization procedure is proposed to obtain a suboptimal solution to the problem (\ref{eq:original1}).
\subsection{Rank Reduction Procedure}
The rank reduction procedure for general separable SDP of form \eqref{eq:conventional_problem2} has been proposed in \cite{huang2010rank, huang2010dual}. By employing a modified stopping criteria, the rank reduction procedure is applied to compute a rank reduced solution from any optimal solution of \eqref{eq:relaxed}.\\
\indent Let $\{\mathbf{X}_i^{\star}\}_{i=1}^{M}$ denote a solution of the SDR problem (\ref{eq:relaxed}).
The rank reduction algorithm successively reduces the rank of $\{\mathbf{X}_i^{\star}\}_{i=1}^{M}$ as follows. Introducing the factorization $\mathbf{X}_i^{\star} \triangleq \mathbf{Q}_i\mathbf{Q}_i^{H}$ where $\text{rank}(\mathbf{X}_i^{\star})=\text{rank}(\mathbf{Q}_i)= K_i$. Starting from the given solution, the algorithm solves the following homogeneous system of equations corresponding to \eqref{eq:mainconstraint}
\begin{equation}
\sum \limits_{m=1}^{M} {\rm Tr}(\mathbf{Q}_m^{H}\mathbf{A}_{lm}\mathbf{Q}_m\mathbf{\Delta}_m) = 0\quad l \in \{1,\dots,M+L\}
\label{eq:upper_limits}
\end{equation}
where $\mathbf{\Delta}_m \in \mathbb{C}^{K_m\times K_m}$ represents an unknown arbitrary Hermitian matrix. The number of real unknowns in \eqref{eq:upper_limits} equals $\sum \limits_{i=1}^{M} \text{rank}^2(\mathbf{X}^{\star}_i)$,  whereas the number of equations in \eqref{eq:upper_limits} is $M+L$. Hence \eqref{eq:upper_limits} must admit a nontrivial solution when the following inequality \cite{huang2010rank, huang2010dual}
\begin{equation}
\sum \limits_{i=1}^{M} \text{rank}^2(\mathbf{X}^{\star}_i) = \sum_{i = i}^M K_i^2 \leq M+L
	\label{eq:inequality}
\end{equation}
 is violated. A solution $\{\mathbf{\tilde{X}}_i^{\star}\}_{i=1}^{M}$ that exhibits a reduced rank can then be computed as
\begin{equation}
\mathbf{\tilde{X}}_i^{^{\star}} = \mathbf{Q}_i(\mathbf{I} - \frac{1}{\delta_{\max}}\mathbf{\Delta}_i)\mathbf{Q}_i^{H}
\label{eq:updated}
\end{equation}
where $\delta_{\max}$ is the largest eigenvalue of all the matrices of $\{\mathbf{\Delta}_i\}_{i=1}^{M}$ that satisfy \eqref{eq:upper_limits}. 
In \cite{huang2010rank}, the above steps are repeated by assigning $\{\mathbf{\tilde{X}}_i^{\star}\}_{i=1}^{M}$ to be a new input of the algorithm until the inequality \eqref{eq:inequality} is fulfilled. However, in our approach the rank reduction procedure is stopped after the maximum number of iterations $\textit{max\_iter}$ is reached. The modified stopping criteria ensures that the ranks of $\{\mathbf{\tilde{X}}_i^{\star}\}_{i=1}^{M}$ cannot be further reduced when $\textit{max\_iter}$ is set as follows
\begin{equation}
\textit{max\_iter}=\sum \limits_{i=1}^{M}\text{rank}(\mathbf{X}^{\star}_i)-M,
\end{equation}
while the ranks still could be reducible if the stopping criteria in \cite{huang2010rank} is employed. Note that for each iteration it requires to solve a homogeneous system of linear equations, matrix decomposition and singular value decomposition. But compared with the optimization problem (\ref {eq:relaxed}), the operation cost of the iterations is comparatively small. The rank reduction procedure is summarized in Algorithm 1.\\
\begin{algorithm}[htb]
\caption{Rank reduction procedure}
\begin{algorithmic}
\STATE \textbf{Input} $\{\mathbf{X}_i^{\star}\}_{i=1}^{M}$ \text{an optimal solution to the problem (\ref{eq:relaxed})},
\STATE $\qquad ~$ $\{\mathbf{A}_{lm}\}_{m=1,\dots,M; l=1,\dots, M+L}$,\\
\STATE $\qquad ~$ $\textit{max\_iter}$ \text{maximum number of iterations;} \\
\STATE \textbf{Output} \text{$\{\mathbf{X}_i^{\star}\}_{i=1}^{M}$ such that the rank of any of the matrices}\\
\STATE $\qquad ~~$ \text{$\{\mathbf{X}_i^{\star}\}_{i=1}^{M}$ cannot be further reduced;}\\
\WHILE{\text{Number of iterations} $\leq$ $\textit{max\_iter}$}
\STATE \text{Decompose}  $\mathbf{X}_i^{\star} = \mathbf{Q}_i\mathbf{Q}_i^{H}$ 
\text{$\quad \forall i=1,\dots,M$;}
\STATE \text{Find a non-zero solution of the equation (\ref{eq:upper_limits});} \\
\IF{\text{(\ref{eq:upper_limits}) does not admit a nontrivial solution}}
	\STATE \textbf{break}
\ELSE
	\STATE \text{Let $\delta_{\max} \triangleq \max_{\substack{1\leq l\leq K_m \\ 1\leq i\leq M}}\{|\delta_{li}|\}$ where $\delta_{li}$ is}\\
	\STATE \text{the $l$-th eigenvalue of $\mathbf{\Delta}_i$;}
	\STATE \text{Set $\mathbf{X}_i^{^{\star}} = \mathbf{Q}_i(\mathbf{I} - \frac{1}{\delta_{\max}}\mathbf{\Delta}_i)\mathbf{Q}_i^{H} \quad \forall i=1,\dots,M$;}\\
\ENDIF
\ENDWHILE
\end{algorithmic}
\end{algorithm}
\indent Next we derive conditions on the number of additional shaping constraints and the code dimension $K$ of the real-valued OSTBC for which optimal beamforming solution can always be obtained. These conditions are stated by the following lemma.
\newtheorem*{lemma2}{Lemma 2}
\begin{lemma2}
Assume that the relaxed problem \textup{(\ref{eq:relaxed})} and its dual \eqref{eq:dual_problem} are solvable\footnote{``solvable'' means that a bounded optimal value of the optimization problem can be obtained \cite{huang2010rank}.} and that the condition
\begin{equation}\label{eq:cond_optimality}
L \leq (K+1)^2 - 2
\end{equation}
is satisfied, then there always exists an optimal solution $\mathbf{X}^{\star}_i$ for problem \textup{(\ref{eq:relaxed})} with $\mathrm{rank}(\mathbf{X}^{\star}_i) \leq K$ for all $i=1,\dots,M$.
\end{lemma2}
\begin{IEEEproof}
We follow a similar line of argument as in \cite{huang2010rank} and prove Lemma 2 by contradiction.   Assume that \eqref{eq:cond_optimality} is satisfied and there exists a matrix $\mathbf{X}_j^{\star}$ with $\mathrm{rank}(\mathbf{X}^{\star}_j)>K$ for some $j$ such that the matrices \footnote{We observe that none of the matrices $\{\mathbf{X}_i^{\star}\}_{i=1}^{M}$ are zero matrices, as otherwise at least one of the $\text{SINR}$ constraints in (\ref{eq:sinr_constraints}) would be violated due to the positive semidefiniteness of $\{\mathbf{X}_i^{\star}\}_{i=1}^{M}$ and the definition of $\{\mathbf{A}_{im}\}_{i,m=1}^{M}$ in (\ref{eq:Amatrix}). Hence all the matrices $\{\mathbf{X}_i^{\star}\}_{i=1}^{M}$ must have a rank greater than or equal to one.} $\{\mathbf{X}_i^{\star}\}_{i=1}^{M}$ satisfy \eqref{eq:inequality}. Then
\begin{equation}
\sum \limits_{i=1}^{M} \text{rank}^2(\mathbf{X}^{\star}_i) \overset{(a)}{\geq} M-1 + (K+1)^2 \overset{(b)}{>} M+L
\label{eq:constraint_inequality}
\end{equation}
where strict equality holds in ``$(a)$'' if and only if there are $M\!-\!1$ rank-one matrices in $\{\mathbf{X}_i^{\star}\}_{i=1}^{M}$ and the last matrix has rank $K+1$, and the strict inequality in ``$(b)$'' follows from (\ref{eq:cond_optimality}). The inequality \eqref{eq:constraint_inequality} however contradicts our assumption that \eqref{eq:inequality} is fulfilled. Hence all the matrices $\{\mathbf{X}_i^{\star}\}_{i=1}^{M}$ must have ranks less than or equal to $K$. We conclude that the maximum number of additional shaping constraints $L$ for which a rank less than or equal to $K$ can be obtained is given by
\begin{equation}
L = (K+1)^2 - 2.
\label{eq:final_ineq}
\end{equation}
\end{IEEEproof}
Lemma 2 indicates that we can always find an optimal solution to problem (\ref{eq:original1}) by using the SDR approach and the rank reduction procedure described in Algorithm 1 if condition \eqref{eq:final_ineq} is satisfied. From \eqref{eq:final_ineq}, we can calculate the maximum numbers of additional shaping constraints for different choices of $K\in\{1,2,4,8\}$ as listed in Table \ref{tab:table_2} such that an optimal solution to problem (\ref{eq:original1}) can always be obtained. We observe from Table \ref{tab:table_2} that our proposed scheme can accommodate a maximum number of 79 additional shaping constraints which corresponds to the choice of the code dimension $K=8$. 

Since a smaller code size of the real-valued OSTBC matrix results in a shorter decoding latency at the receiver side, we seek to obtain the smallest value of $K$ for all downlink users based on the output of the rank reduction procedure in Algorithm 1. If the updated $\{\textbf{X}_i^{\star}\}_{i=1}^M$ after the rank reduction procedure satisfies ${\rm rank}(\textbf{X}_i^{\star})\leq8$ for all $i=1,\dots,M$, then the smallest number $K$ is chosen from $K\in\{1,2,4,8\}$ such that
\begin{align}
	K \geq {\rm rank}(\textbf{X}_i^{\star})~~ \forall i=1,\dots,M.
\end{align}
The corresponding beamforming matrices are then obtained as
\begin{equation}
\label{Bmatrices}
\textbf{W}_i^{\star} = [\textbf{Q}_i,\, \mathbf{0}_{N\times (K-{\rm rank}(\textbf{X}_i^{\star})}]
\end{equation}
where $\mathbf{X}_i^{\star} = \mathbf{Q}_i\mathbf{Q}_i^{H}$ with $\mathbf{Q}_i \in \mathbb{C}^{N \times {\rm rank}({\mathbf{X}_i^{\star}})}$.

\subsection{General Rank Randomization Procedure}
In the case that (\ref{eq:relaxed}) is feasible, \eqref{eq:cond_optimality} is violated and if at least one of the matrices in $\{\mathbf{X}_i^{\star}\}_{i=1}^{M}$, after rank reduction, exhibits a rank greater than eight, the following randomization technique which involves a power control problem could be applied to generate a feasible but generally suboptimal beamforming solution for problem \eqref{eq:original1}. Note that the randomization procedure may lead to an infeasible solution if the power control problem is infeasible. Note that the randomization procedure may not find a feasible suboptimal solution if the power control problem for each randomization sample is infeasible.

Let us decompose the matrices $\{\mathbf{X}_i^{\star}\}_{i=1}^{M}$ into
\begin{equation}
 \mathbf{X}_i^{\star} = \mathbf{U}_i\boldsymbol{\Sigma}_i\mathbf{U}_i^{H}.
\label{eq:eig_decomposition}
\end{equation}
The corresponding beamforming matrices $\{\mathbf{\bar{W}}_i\}_{i=1}^{M}$ are then randomly generated according to
\begin{equation}
 \bar{\mathbf{W}}_i \triangleq \left[\mathbf{\bar{w}}_{i1}, \mathbf{\bar{w}}_{i2}, \ldots, \mathbf{\bar{w}}_{i8} \right]= \mathbf{U}_i\boldsymbol{\Sigma}_i^{1/2}\boldsymbol{\Lambda}_i
\label{eq:randomization}
\end{equation}
where $\boldsymbol{\Lambda}_i$ is the $N\times 8$ matrix whose elements are drawn from an i.i.d. complex circular Gaussian distribution with zero mean and unit variance. Note that the instances of the beamforming matrices $\{\mathbf{\bar{W}}_i\}_{i=1}^{M}$ generated in \eqref{eq:randomization} are generally not feasible for problem \eqref{eq:original1}. In order to compute a feasible solution with spatial characteristics corresponding to $\{\mathbf{\bar{W}}_i\}_{i=1}^{M}$, a power control problem involving linear programming is solved. The randomization procedure is summarized in Algorithm 2.


With the rank reduction procedure in Algorithm 1 and the randomization procedure in Algorithm 2, a solution to problem \eqref{eq:original1} can be computed following the procedure summarized in Algorithm 3.
\begin{algorithm}
\caption{Randomization procedure}
\begin{algorithmic}
\STATE \textbf{Input} \text{$\{\mathbf{X}_i^{\star}\}_{i=1}^{M}$ with ranks greater than 8 for some $i$,}
\STATE $\qquad ~$ \text{$N_{\text{rand}}$ number of iterations,}
\STATE $\qquad ~$ \text{$P_{\text{opt}}$ optimal value of the power control problem;}
\STATE \textbf{Output} \text{The scaled beamforming matrices}
\STATE $\qquad \quad$ \text{of the problem \eqref{eq:relaxed};}
\STATE \text{Set $K=8$, \text{$P_{\text{opt}} = +\text{Inf}$;}}
\FOR{$k=1$ to $N_{\text{rand}}$}
	\STATE \text{Obtain $\mathbf{\bar{W}}_i$ according to (\ref{eq:randomization});}
	\STATE \text{Solve the power control problem;}
	\IF {\text{The optimal value is less than $P_{\text{opt}}$}}
		\STATE \text{Set $P_{\text{opt}}$ to be equal to the optimal value and}
		\STATE \text{store the scaled beamforming matrices}
	\ELSE
	\STATE \text{Discard the matrices $\{\mathbf{\bar{W}}_i\}_{i=1}^{M}$;}
	\ENDIF
\ENDFOR
\end{algorithmic}
\end{algorithm}
\begin{table}%
\centering
\resizebox{0.8\columnwidth}{!}{
\begin{tabular}{|c|c|}
\hline
Number of beamformers  & Number of additional\\
per user $K$& shaping constraints \\
\hline
1&2\\
\hline
2& 7\\
\hline
4& 23\\
\hline
8& 79\\
\hline
\end{tabular}}
\caption{Number of additional shaping constraints}
\label{tab:table_2}
\end{table}

\begin{algorithm}
\label{algsummary}
\caption{Summary}
\begin{algorithmic}
\STATE \textbf{Input} \text{$\{\mathbf{X}_i^{\star}\}_{i=1}^{M}$  an optimal solution to the problem (\ref{eq:relaxed});}
\STATE \textbf{Output} \text{$\{\mathbf{W}_i^{\star}\}_{i=1}^{M}$ beamforming matrices of the problem}
\STATE $\qquad \quad ~$\text{(\ref{eq:original1}), $K$ number of beamformers per user;}
\IF{$\text{rank}(\mathbf{X}_i^{*}) >1$ for some $i$}
\STATE \text{Apply Algorithm 1 to obtain the rank-reduced}
\STATE \text{matrices $\{\mathbf{X}_i^{\star}\}_{i=1}^{M}$;}
\ENDIF
\IF{$\text{rank}(\mathbf{X}_i^{\star})\leq 8~ \forall i=1,\dots,M$}
\STATE \text{Choose $K$ to be the smallest number out of $\{1,2,4,8\}$}
\STATE \text{such that $\text{rank}(\mathbf{X}_i^{*}) \leq K\quad \forall i=1,\dots,M$;}
\STATE \text{Decompose $\{\mathbf{X}_i^{\star}\}_{i=1}^{M}$ to obtain $\{\mathbf{W}_i^{\star}\}_{i=1}^{M}$ using \eqref{Bmatrices}};\\
\ELSE{}
\STATE \text{Apply Algorithm 2 to obtain suboptimal}
\STATE \text{beamforming matrices $\{\mathbf{W}_i^{\star}\}_{i=1}^{M}$;}
\ENDIF
 \STATE \text{Rotate matrices $\{\mathbf{W}_i^{\star}\}_{i=1}^{M}$ if necessary according to (\ref{phaseconstraints1}).}
\end{algorithmic}
\end{algorithm}
\section{Simulations}
Four simulation examples are provided to demonstrate the performance of our proposed downlink beamforming scheme with a large number of additional shaping constraints of different types. We assume that the base station is equipped with a uniform linear array (ULA), and the transmit antennas are spaced half wavelength apart. According to our system model, accurate CSI for all users and terminals is available at the transmitter. The noise powers of the downlink users in all examples are assumed to be $\sigma_i^2=0.1$ for $i = 1, ..., M$. We also declare that $\text{rank}(\mathbf{X}_m^{\star})= \xi$ if the $(\xi\!+\!1)$-th largest eigenvalue is smaller than $0.01\%$ of the sum of all eigenvalues.

\subsection{Example 1}
In the first example, we consider the design of downlink beamformers with external wireless charging terminals. Considering a line-of-sight transmission scenario, we assume that three downlink users $(M=3)$ connected to the base station are located at directions $\theta_1=-5^{\circ}$, $\theta_2=10^{\circ}$ and $\theta_3=25^{\circ}$ relative to the array broadside of the serving base station. The number of antennas at the base station is $12$ $(N=12)$. We assume that there are $22$ charging terminals, which are centered around
\begin{align}\vartheta_{4,\dots,14}=[&-80^{\circ},-75^{\circ},-70^{\circ},-65^{\circ},-60^{\circ},-55^{\circ},\nonumber\\&-45^{\circ},-35^{\circ},-25^{\circ},-8^{\circ},-2^{\circ} ] \label{eq:chancelpos}
\end{align} and
 \begin{align}
\vartheta_{15,\dots,25}=[&12^{\circ},18^{\circ},35^{\circ}, 45^{\circ}, 50^{\circ}, 55^{\circ},\nonumber\\& 60^{\circ}, 65^{\circ}, 70^{\circ}, 75^{\circ}, 80^{\circ}] \label{eq:chancelneg}
\end{align}
relative to the serving base station of the cell under consideration. For all downlink users and charging terminals, the spatial signatures are modeled as
\begin{eqnarray}
\mathbf{h}(\theta_m) \! &=\!&  \begin{bmatrix}1, e^{j\pi\sin(\theta_m)}, \dots, e^{j\pi(N-1)\sin(\theta_m)}\end{bmatrix}^{T}\nonumber\\
&& m \in \{1,\dots,25\},
\label{eq:channels}
\end{eqnarray}
i.e., the path loss of all downlink users and charging terminals is identical \cite{huang2010rank}. To make our simulation results more meaningful, we randomly vary the locations of the downlink users and the charging terminals in different Monte-Carlo runs, i.e., the angles of departure at the base station are simulated as
\begin{eqnarray}
\theta_m &= \vartheta_m + \Delta \theta_m \quad m \in \{1,\dots,25\},
\label{eq:vartheta}
\end{eqnarray}
where the random variations $\Delta\theta_m$ drawn from a uniformly distributed within the interval $[-0.25^{\circ}, 0.25^{\circ}]$. We use the additional shaping constraints in (\ref{eq:mainconstraint}) to ensure predefined charging power levels at the $l$-th charging terminal in each time slot where $\mathbf{A}_{lm} = \mathbf{h}(\theta_l)\mathbf{h}^{H}(\theta_l)$ for $m=1,...,M$ and $l=4,...,25$. We set the minimum power threshold $b_l$ to be $5$dB for each charging terminal and $\unrhd_l = \; \geq$. The SINR targets $\gamma_i$ at the individual downlink users are varied between $0$dB and $10$dB. The simulation results are averaged over 300 Monte-Carlo runs. In each run, the number of randomization instances is set to $300$ for all approaches if necessary.

\begin{figure}%
\centering
\includegraphics[width=\columnwidth]{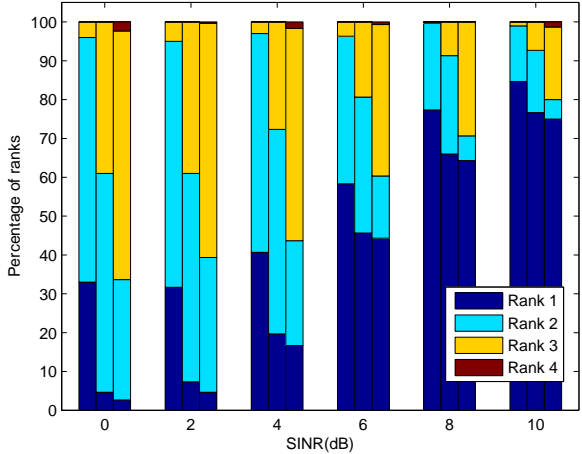}%
\caption{The ranks of the matrices $\mathbf{X}_1^{\star}$ (left bar), $\mathbf{X}_2^{\star}$ (middle bar), $\mathbf{X}_3^{\star}$ (right bar) after the rank reduction procedure.}%
\label{fig:RankPercentage_ex1}%
\end{figure}

\begin{figure}%
\centering
\includegraphics[width=\columnwidth]{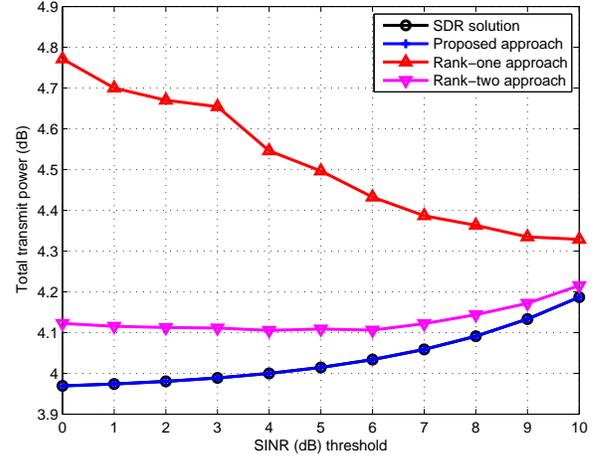}%
\caption{Transmitted power per time slot at the base station.}%
\label{fig:PowervsSINR}%
\end{figure}

The ranks of the solution matrices of the relaxed problem (\ref{eq:relaxed}) after the rank reduction procedure are plotted in Fig. \ref{fig:RankPercentage_ex1}. According to the reduced rank property provided in Table \ref{tab:table_1}, the code dimension is selected as $K = 4$.
It can be analytically proven from a power scaling argument that problem \eqref{eq:relaxed} , in the case of power charging constraints, is always feasible for all approaches. In Fig. \ref{fig:PowervsSINR} we display the total transmitted power per time slot at the base station versus the SINR  for different approaches. As shown in Fig. \ref{fig:PowervsSINR}, the proposed general rank beamforming approach outperforms the competing approaches in terms of transmitted power. In the low SINR region, the gap between the rank-one and rank-two approaches and the proposed approach is large, because as shown in Fig. \ref{fig:RankPercentage_ex1}, most of the solution matrices are of high rank ($\geq\!\!2$) and thus the suboptimal randomization approximation is performed in the rank-one and rank-two approaches. In the high SINR region, the gaps between different approaches decrease because as shown in Fig. \ref{fig:RankPercentage_ex1}, the percentage of rank-one solution matrices increases which results in an increased number of optimal solutions for all approaches.

\subsection{Example 2}
In the second example, we consider the downlink beamformer design according to problem (\ref{eq:relaxed}) for beam pattern (BP) with smooth and flat sidelobes to reduce the interference to co-channel users. We assume that in our simulation scenario the base station consists of $18$ antennas $(N=18)$. In this simulation the locations of three downlink users $(M=3)$ are the same as in the previous example, i.e., $\theta_1=-5^{\circ}$, $\theta_2=10^{\circ}$ and $\theta_3=25^{\circ}$. The SINR thresholds of downlink users are set to $\gamma_i=10$dB. Moreover, we assume that nineteen co-channel users connected to a neighboring base station are present in the scenario, which are located at
\begin{align}
&\mu_{1,\dots,19} = [-89.375^{\circ}, -80^{\circ}, -70.625^{\circ}, -61.25^{\circ}, -51.875^{\circ}, \nonumber \\& 	-42.5^{\circ}, -33.125^{\circ}, -23.75^{\circ}, -14.375^{\circ}, 2^{\circ}, 3^{\circ}, 17^{\circ},  18^{\circ}, \nonumber \\& 34.375^{\circ}, 43.75^{\circ}, 53.125^{\circ}, 62.5^{\circ}, 71.875^{\circ}, 81.25^{\circ}].
\end{align}
The channel propagation model is the same as defined in (\ref{eq:channels}). The interference power at the direction $\mu_l$ relative to the base station in each time slot can be written as
\begin{equation}
f(\mu_j) = \sum \limits_{m=1}^{M}{\rm Tr}(\mathbf{A}_{(j+3)m}\mathbf{X}_m)
\end{equation}
where $\mathbf{A}_{(j+3)m} = \mathbf{h}(\mu_{j})\mathbf{h}^{H}(\mu_{j})$ for $m=1,...,M$ and $j=1,...,19$.
In our beamformer design, the interference power is upper bounded by $b_{j+3}=0.1$ and $\unrhd_{j+3} = \; \leq$ for $j=1,...,19$. In addition to these constraints, we guarantee that the interference power at the direction $\mu_l$ attains a local minimum value by adding interference derivative constraints, i.e., the interference in the vicinity of the constraint directions remains approximately constant if
\begin{eqnarray}
&& -\epsilon_a \leq \frac{df(\mu_j)}{d\mu_j} \leq \epsilon_a \quad \text{and} \quad \frac{d^2f(\mu_j)}{d\mu_j^2} >0\nonumber\\
&& \quad j \in \{1,\dots,19\}
\end{eqnarray}
where the threshold is set to $\epsilon_a = 10^{-5}$,
\begin{eqnarray}
\frac{df(\mu_{\bar j})}{d\mu_{\bar j}}= \sum \limits_{m=1}^{M}{\rm Tr}(\mathbf{A}_{(j+3)m}\mathbf{X}_m), \quad j \in \{20,\dots,38\}
\end{eqnarray}
and
\begin{eqnarray}
\frac{d^2f(\mu_{\bar j})}{d\mu_{\bar j}^2}=\sum \limits_{m=1}^{M} {\rm Tr}(\mathbf{A}_{(j+3)m}\mathbf{X}_m), \quad j \in \{58,\dots,76\}
\end{eqnarray}
are satisfied, for $m=1,...,M$,
\begin{align}
\mathbf{A}_{({j+3})m}\!&=\!
\begin{cases}
 \!\frac{d\mathbf{h}(\mu_{\bar j})}{d\mu_{\bar j}}\mathbf{h}^{H}(\mu_{\bar j}) \!+\! \mathbf{h}(\mu_{\bar j}){\frac{d\mathbf{h}^{H}(\mu_{\bar j})}{d\mu_{\bar j}}}, \!\!& \!\!j=20,...,38 \\
 \!\frac{d\mathbf{h}(\mu_{\bar j})}{d\mu_{\bar j}}\mathbf{h}^{H}(\mu_{\bar j}) \!+\! \mathbf{h}(\mu_{\bar j}){\frac{d\mathbf{h}^{H}(\mu_{\bar j})}{d\mu_{\bar j}}}, \!\!& \!\!j=39,...,57 \\
  \!\mathbf{h}(\mu_{\bar j}){\frac{d^2\mathbf{h}^{H}(\mu_{\bar j})}{d\mu_{\bar j}^2}}\!+\! \frac{d^2\mathbf{h}(\mu_{\bar j})}{d\mu_{\bar j}^2}{\mathbf{h}^{H}(\mu_{\bar j})} + \!\!& \\
  \!2\frac{d\mathbf{h}(\mu_{\bar j})}{d\mu_{\bar j}}{\frac{d\mathbf{h}^{H}(\mu_{\bar j})}{d\mu_{\bar j}}},
   \!\!& \!\!j=58,...,76
  \end{cases} \\
b_{j+3}&=
\begin{cases}
\epsilon_a, & j=20,...,38 \\
 -\epsilon_a, & j=39,...,57 \\
0,  & j=58,...,76
  \end{cases} \\
 \unrhd_{j+3}&=
  \begin{cases}
  \leq, & j=20,...,38 \\
   \geq, & j=39,...,57 \\
  \geq,  & j=58,...,76
    \end{cases}
\end{align}
with $\bar j \triangleq j \mod{19}$, i.e., the remainder of $j$ divided by $19$. The received sum power at direction $\theta$ relative to the base station, referred to as the sum BP, is defined as
\begin{equation}
\sum \limits_{m=1}^{M}\|\mathbf{h}(\theta)\textbf{W}_m^{\star}\|_2^2
\end{equation}
where $\textbf{W}_m^{\star}$ is the rank reduced solution given in (\ref{Bmatrices}).

\begin{figure}
\includegraphics[width=\columnwidth]{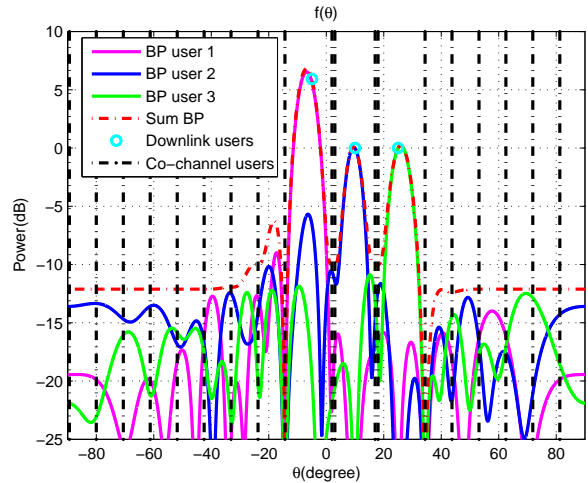}%
\caption{User BPs and sum BP with smoothed and suppressed sidelobes.}%
\label{fig:figure_1}%
\end{figure}

The BPs are presented in Fig. \ref{fig:figure_1}, and the rank properties of the solution matrices are provided in Table \ref{tab:table_1}.
Note that there is a total number of $76$ additional shaping constraints in this simulation. According to Lemma $2$, we can find an optimal solution to the optimization problem (\ref{eq:relaxed}) with the rank less than or equal to $8$ by using the rank reduction procedure. Based on the results in Table \ref{tab:table_1}, we select the code dimension $K = 4$.
\begin{table}%
\centering
\resizebox{0.6\columnwidth}{!}{
\begin{tabular}{|c|c|c|c|}
\hline
& $\mathbf{X}_1^{\star}$& $\mathbf{X}_2^{\star}$&$\mathbf{X}_3^{\star}$\\
\hline
Original rank in (\ref{eq:relaxed})&14&15&15\\
\hline
Reduced rank&2&3&4\\
\hline
\end{tabular}}
\caption{Rank property before and after applying rank reduction algorithm.}
\label{tab:table_1}
\end{table}
As shown in Fig. \ref{fig:figure_1}, the proposed approach is capable of coping with a large number of additional shaping constraints.
Furthermore, as listed in Table \ref{tab:table_1}, the ranks of the solution matrices have been significantly reduced which demonstrate the effectiveness of the rank reduction procedure.

\subsection{Example 3}
\begin{figure}%
\centering
\includegraphics[width=\columnwidth]{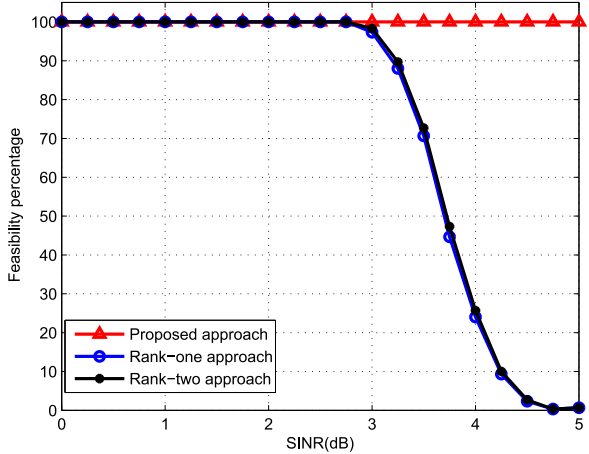}%
\caption{The feasibility percentage of all approaches.}%
\label{fig:feasibility}%
\end{figure}

\begin{figure}%
\centering
\includegraphics[width=\columnwidth]{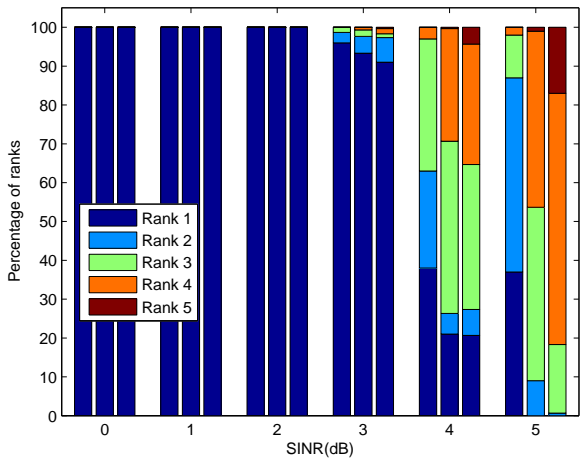}%
\caption{The ranks of the matrices $\mathbf{X}_1^{\star}$ (left bar), $\mathbf{X}_2^{\star}$ (middle bar), $\mathbf{X}_3^{\star}$ (right bar) after the rank reduction procedure.}%
\label{fig:average_rank}%
\end{figure}

The same scenario as in Example $2$ is considered to perform a comparison between our proposed approach with the conventional rank-one and rank-two approaches. All location parameters remain unchanged.
Furthermore, we assume that all angles of departures are also subject to variations in different Monte-Carlo runs, which are defined in the same way as in Example 1.
The required SINRs $\gamma_i$ at the downlink users  are uniformly varied between $0$dB and $5$dB. The results are averaged over 300 independent Monte-Carlo runs and the number of randomization instances in each run is set to $100$ for all approaches if necessary. The feasibility percentage of all approaches is displayed in Fig. \ref{fig:feasibility}. From  Fig. \ref{fig:feasibility}, we observe that the proposed approach is always feasible for different SINR thresholds. In contrast to this, the feasibility of the  rank-one and rank-two approaches decreases with increasing SINR thresholds. This demonstrates that our proposed approach has a wider feasibility range compared to existing approaches.
The ranks of the solution matrices of the relaxed problem (\ref{eq:relaxed}) after the rank reduction procedure are plotted in Fig. \ref{fig:average_rank}.
As shown in Fig. 4, when $\gamma_i<3$dB, all three approaches are feasible. This is due to the fact that in this case, as shown in Fig. 5, rank-one solutions are obtained for all approaches. In other words, optimal solutions are obtained for all approaches and thus the performance obtained from all approaches is identical. Therefore, when $\gamma_i<3$dB, the code dimension for our proposed method is chosen as $K=1$. In contrast to this when $\gamma_i\geq3$dB we observe from Fig. 5 that the rank of the optimal solutions takes different values in the range between one and five. Thus in contrast to the rank-one and rank-two beamforming approaches if a rank larger than two is obtained, our proposed approach retains the optimality property and yields feasible solutions while the competing approaches yield suboptimal solutions or even become infeasible for $\gamma_i\geq3$dB.
\newline
\subsection{Example 4}
\begin{figure}%
\includegraphics[width=\columnwidth]{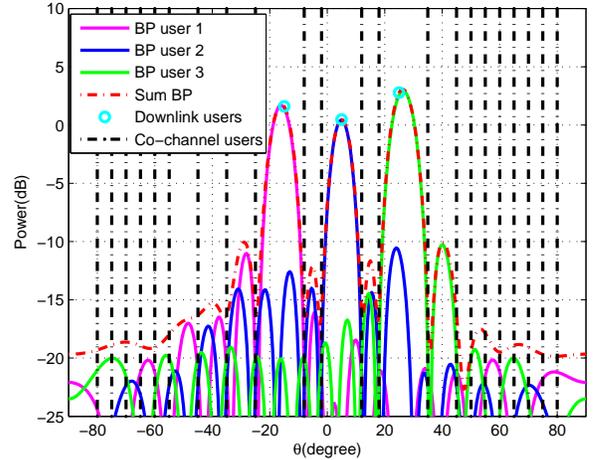}%
\caption{User BPs and sum BP of downlink beamforming problem subject to interference power constraints.}%
\label{fig:figure_2}%
\end{figure}

The aim of the fourth example is to demonstrate the interference power suppression at each co-channel user to a fraction of its maximum value. In this example the concept of relaxed nulling is used to formulate the additional (indefinite) shaping constraints for interference power limitation \cite{hammarwall2006}. The base station under consideration is equipped with a ULA of 15 antennas that are spaced half wavelength apart $(N=15)$. Three downlink users served by the base station are located at  $\theta_1=-15^{\circ}$, $\theta_2=5^{\circ}$ and $\theta_3=25^{\circ}$ relative to the base station. We assume that twenty two co-channel users served by neighboring base stations are present in our scenario which are located at the same position as in (\ref{eq:chancelpos}) and (\ref{eq:chancelneg}).
We set the SINR thresholds to the same value as in Example $2$. Similarly, the spatial signatures are modeled according to (\ref{eq:channels}). We limit the interference power to the coexisting users by the following constraints
\begin{eqnarray}
&&{\rm Tr}(\mathbf{h}(\theta_j)\mathbf{h}(\theta_j)^{H}\mathbf{X}_i) \leq \beta\|\mathbf{h}(\theta_j)\|_2^2{\rm Tr}(\mathbf{X}_i)\nonumber\\
&& \forall i=1,2,3, \quad \forall j=4,\dots,25
\end{eqnarray}
where $\beta \ll 1$ is an interference constraint parameter.
The above constraints can be reformulated into the form of (\ref {eq:mainconstraint}) where, for $\tilde m, m=1,...,3$,
\begin{align}
\mathbf{A}_{(22(\tilde m -1)+j)m}\!&=\!
\begin{cases}
\beta\|\mathbf{h}(\theta_j)\|_2^2\mathbf{I}  - \mathbf{h}(\theta_j)\mathbf{h}(\theta_j)^{H}, \!\!& \!\!
 \tilde m = m \\
  \mathbf{0}, \!\!& \!\!\tilde m \not = m
  \end{cases} \\
b_{n+3}&= 0, \\
 \unrhd_{n+3}&=\geq, \quad n=1,...,66;\ j=4,...,25.
\end{align}
We note that the matrix $\mathbf{A}_{lm}$ is either zero or indefinite for $l=4,...,69, m=1,...,3$ and there is a total number of $66$ additional shaping constraints in this simulation. In the simulation, $\beta$ is chosen to be $0.5\%$. As shown in Fig. \ref{fig:figure_2}, the interference power at the locations of the coexisting users is limited to a reasonable level.
\section{Conclusion and outlook}
In the paper, we propose a general rank beamforming approach for the multiuser downlink beamforming problem with additional shaping constraints. The general rank approach increases the degrees of freedom in the beamformer design by using high dimensional full-rate real-valued OSTBC. In our proposed approach, an optimal solution can be obtained when the ranks of all SDR solution matrices are less than or equal to eight after the rank reduction procedure. Moreover, in our scheme an optimal solution for the original problem can be found when the number of additional shaping constraints is less than or equal to $79$. The range of applications for our proposed beamforming scheme is hence much wider than that of the conventional rank-one and rank-two approaches. Our proposed general rank beamforming framework exhibits an underlying optimization problem structure that is similar to that of the conventional rank-one and rank-two beamforming approaches. This allows a simple extension of the approach to existing robust beamforming designs, e.g., in the practically important case of inaccurate CSI, maintaining the benefits of increased degrees of freedom available in the proposed general rank beamforming approach. All the results presented in this paper can be extended to QCQP problems with double-sided constraints considered in \cite{Newwork} as an interesting topic of future research.
\section{Acknowledgement}
This work was supported by the Seventh Framework Programme for Research of the European Commission under grant number ADEL-619647.
\nocite{*}
\bibliographystyle{IEEEtran}	
\bibliography{references}

\end{document}